\documentclass[
aps,
pre,
reprint,
amsmath,
amssymb,
longbibliography,
superscriptaddress
]{revtex4-2}

\usepackage{graphicx}
\usepackage{dcolumn}
\usepackage{bm}
\usepackage{amsmath}
\usepackage{amssymb}
\usepackage{physics}
\usepackage{color}
\usepackage{subfigure}
\usepackage{booktabs}
\usepackage{hyperref}
\usepackage{placeins}

\begin{document}

\title{Improved  Finite-Time Lyapunov Exponent framework for Comprehensive Synchronization Analysis in Multilayer Networks}

\author{R.~Vigneshwaran and N.Athavan}
\email{n.athavan@hhrc.ac.in}
\affiliation{Department of Physics, H.H.~The Rajah's College (Autonomous and Affiliated to Bharathidasan University, Tiruchirappalli 620024, Tamil Nadu, India), Pudukkottai 622001, Tamil Nadu, India}

\begin{abstract}
 We propose an Improved version of the Lagrangian Finite Time Lyapunov Exponents(IFTLE) framework as a new quantifier for Comprehensive Synchronization analysis of multilayer networks undergo adaptive coupling scheme in order to incorporate the effects of deformation in synchrony due to the dynamic variations in the coupling strengths of intra-layer and inter-layer links. Theoretical formulation of the IFTLE along with its numerical validation has been provided using a three layer network of Chaotic R\"ossler oscillators. The numerical illustration clearly shows the efficacy of the proposed method  revealing the need for comprehensive synchrony deformation analysis in adaptive coupling scenarios. Also we have stated a few properties of IFTLE useful for the quantification purpose.  
 \end{abstract}

\maketitle

\section{Introduction}
Synchronization is one of the most celebrated Collective phenomena in the realm of complex networks~\cite{ref1,ref2}. These networks range from power grids, neuronal circuits, cyber-physical infrastructures, and so on are frequently organized into multiple interacting layers whose collective behaviors, including chimera states, relay synchronization, and partial synchronization, depend jointly on intralayer and interlayer coupling configurations~\cite{ref20,ref21,ref22,ref32,ref33,ref34}.The robustness and dependence of chimera states on initial conditions have
also been investigated using basin-stability approaches~\cite{ref35}.
Chimera and multichimera states have also been reported in networks of
bursting neurons under different coupling configurations~\cite{ref36}. Interestingly, in the recent past, synchronization of chaotic oscillators assembled in multilayer networks has been extensively studied~\cite{ref3,ref4}. A bunch of qualitative and quantitative tools are available in the literature to characterize such collective phenomena. Among them, classical Lagrangian Finite Time Lyapunov Exponent(FTLE) which gives the divergence measure of nearby trajectories in phase space, is one of the most reliable quantitative tools to study the underlying dynamics of nonlinear dynamical systems~\cite{ref7,ref8,ref9,ref10,ref11}. However, in certain networks which undergo state dependent time varying coupling scheme, or in other words, adaptive coupling scheme, deformation of the synchronization states takes place every moment altering the course of underlying dynamics. Therefore, a need arises to take care of the instantaneous effect of deformed synchronization states  for studying the comprehensive underlying dynamics of the complete network. But the FTLE framework currently in use is formulated to study asymptotic behavior with respect to single-layer and do not encapsulate the additional state-space deformation produced by adaptive coupling across layers~\cite{ref5,ref6}. Moreover, majority of the existing synchronization criteria are stipulated in terms of asymptotic synchronization behavior without  taking into account the effects of transient synchronization ~\cite{ref12,ref13,ref14}. Also, the fact that FTLE definitions deal only with finite-time trajectory divergence ignoring synchronization-linked deformation arising from interactions within and between network layers~\cite{ref15,ref16,ref17,ref18,ref19} emphasizes to look for due incorporation of adaptive factors to make it a composite quantifier. 

Having the necessity for the reparation of the FTLE frame work in question, and taking it as our primary motivation and objective, we propose an improved version of the FTLE framework which treats the deformation dynamics inclusively in the trajectory divergence analysis in addition to the current Lagrangian framework of FTLE.

 Our proposed framework essentially has three key stages. First, formulation of an Improved Finite-Time Lyapunov Exponent(IFTLE) framework for comprehensive synchronization analysis in multilayer networks with adaptive coupling topology. Second, establishment of connection between IFTLE evolution and the stability of synchronization manifold through analytical and numerical investigations. Third, articulation of synchronization transitions and characterization of incoherent, chimera, partially synchronized, and completely synchronized states pertained to adaptive coupling.

The structure of this paper is as follows. In Section~II we present the general mathematical model of a three layer multilayer network comprising the state-space formulation, the multilayer architecture, the compact network dynamical equations and the adaptive coupling scheme. Section~III deals with the proposed Improved Finite-Time Lyapunov Exponent framework, including the deformation field, the Deformation Tensor, the IFTLE measure, and its fundamental theoretical properties. In Section~IV we undertake the analytical stability analysis of the synchronization manifold in which the error dynamics, the Lyapunov stability analysis, and the IFTLE stability condition are exposed. Section~V describes the numerical validation with the aid of a three-layer R\"ossler oscillator network  and presents the simulation procedure together with the noise perturbation model. Section~VI presents the numerical results and discussion, including synchronization indicators,  IFTLE evolution, comparison with the classical FTLE, and statistical validation. Section~VII discusses the physical interpretation and potential applications of the proposed framework, and Section~VIII encompasses the future direction and a formal conclusion.
\section{Model}

\subsection{Network Architecture}
\begin{figure}[t]
    \centering
    \includegraphics[width=0.5\linewidth]{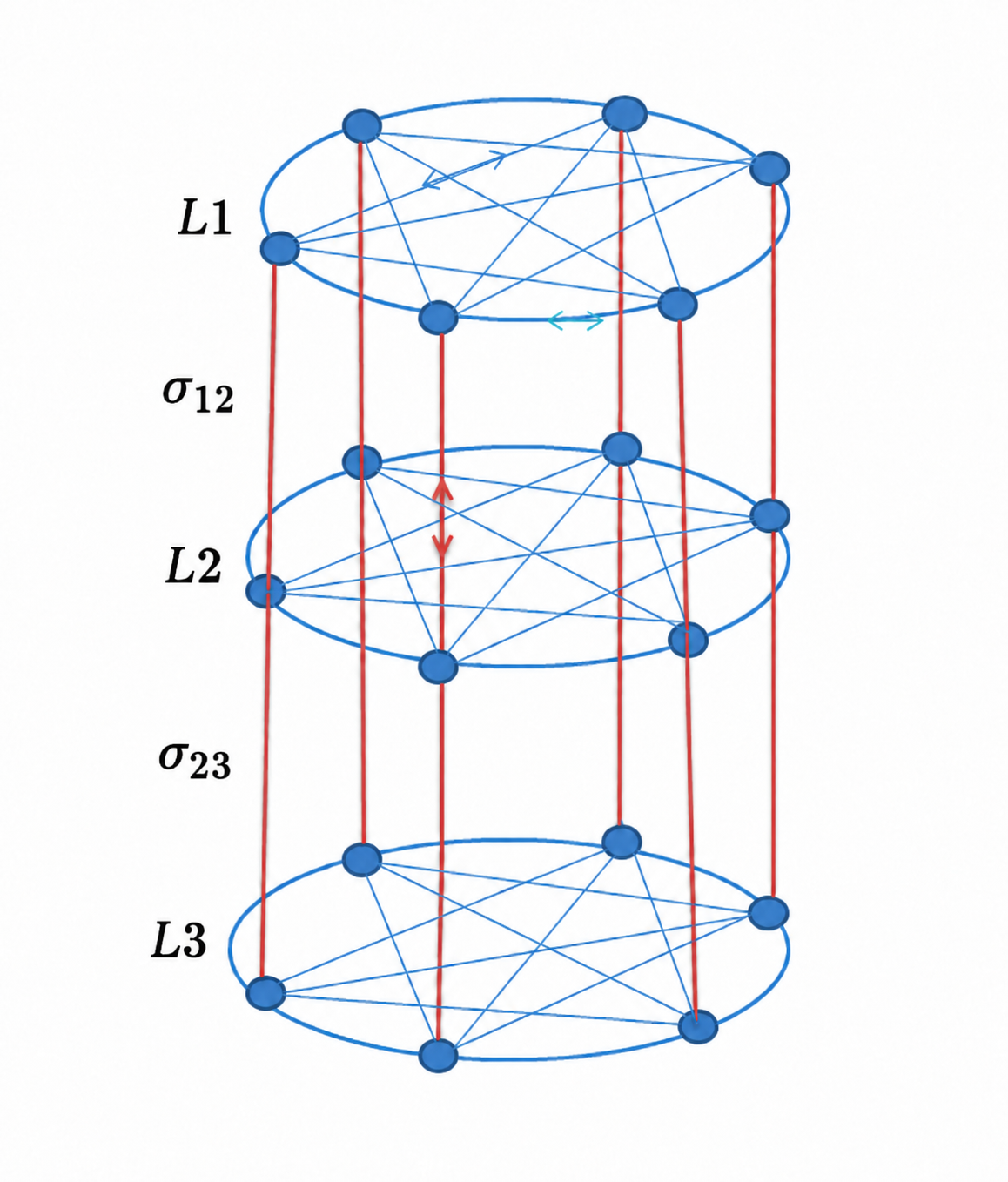}
    \caption{Schematic representation of the considered multilayer network and coupling architecture}
    \label{fig:schematic}
\end{figure}

We consider a multilayer network stacked with three layers whose schematic is given in Fig~\ref{fig:schematic}. The layers at the top and bottom act as outer driver layers and the middle layer, being the relay layer, mediates the interaction between them. Each layer, denoted by $l$, has $N$ oscillators, in the form of a ring, that execute chaotic oscillations and are coupled, to every other oscillator in the layer, under the Global coupling scheme with coupling strength $\sigma$. In addition, every oscillator in a particular layer has interlayer connections to their counterparts in the other layers. Blue colour lines within each layer represent the intralayer global coupling connections, whereas the red vertical lines connect corresponding node indices across adjacent layers and represent the interlayer adaptive couplings facilitating the exchange of synchronization information between neighboring layers. The interlayer coupling strength between the adjacent layer pairs 1 and 2, and 2 and 3  are determined following an adaptive coupling scheme. These interlayer coupling strengths are crucial in the sense that they decide the rate of exchange of state information across layers and are responsible for driving the network toward synchrony as intralayer coupling strengths facilitate the synchronization of the oscillators of the corresponding layer. 

\subsection{Network Equations}

The global state of a general multilayer system is denoted by

\begin{equation}
\mathbf{X}(t)
=
\left\{
\mathbf{X}_{i}^{(\ell)}(t)
\right\},
\qquad
i=1,2,\ldots,N,
\qquad
\ell=1,2,3,
\label{eq:global_state}
\end{equation}
where $\mathbf{X}_{i}^{(\ell)}(t)$ represents the state vector of the
$i$th oscillator in layer $\ell$,and $N$ denotes the number of oscillators in each layer.
The global state vector of our model containing three layers, can be expressed as 

\begin{equation}
\mathbf{X}
=
\begin{bmatrix}
\mathbf{X}^{(1)}\\
\mathbf{X}^{(2)}\\
\mathbf{X}^{(3)}
\end{bmatrix},
\label{eq:global_vector}
\end{equation}

where $\mathbf{X}^{(\ell)}$ denotes the collective state vector of all oscillators in layer $\ell$. We subject our interlayer links of our network to an adaptive coupling scheme in which the momentary change in the underlying dynamics is mainly caused by the instantaneous value of the state vectors.  The dynamics of oscillator $i$ in layer $\ell$ is governed by

\begin{equation}
\begin{aligned}
\frac{d\mathbf{X}_{i}^{(\ell)}}{dt}
={}&
\mathbf{F}\!\left(\mathbf{X}_{i}^{(\ell)}\right)
-
\sigma_{\mathrm{intra}}
\sum_{j=1}^{N}
L_{ij}^{(\ell)}
\mathbf{X}_{j}^{(\ell)}
\\
&
-
\Gamma_i^{(\ell m)}(t)[W_i(t)]_{\ell m}
\left(
\mathbf{X}_{i}^{(\ell)}
-
\mathbf{X}_{i}^{(m)}
\right).
\end{aligned}
\label{eq:network_dynamics1}
\end{equation}
where $\mathbf{F}(\mathbf{X}_{i}^{(\ell)})$ represents the intrinsic
nonlinear dynamics of oscillator $i$ in layer $\ell$,
$L_{ij}^{(\ell)}$ is the $(i,j)$th element of the intralayer
Laplacian of layer $\ell$~\cite{ref31}, and $\sigma_{\mathrm{intra}}$ is the
intralayer coupling strength. The index $m$ denotes a layer adjacent
to layer $\ell$.  $[W_i(t)]_{\ell m}$ contains both the baseline interlayer
coupling strength($\sigma_{12}$ and $\sigma_{23}$) and its time-dependent adaptive weight ($\omega^{(\ell m)}_{i}$ for $\ell$ to $m$) which is written as

\begin{equation}
[W_i(t)]_{\ell m}
=
\sigma_{\ell m}\omega_i^{(\ell m)}(t).
\label{eq:effective_interlayer_weights}
\end{equation}
where $\omega_i^{(\ell m)}(t)$ evolves according to
\begin{equation}
\frac{d\omega^{(\ell m)}_{i}(t)}{dt}
=
-\varepsilon
\left[
\sin\left(\theta^{(\ell)}_i-\theta^{(m)}_i+\gamma\right)
+
\omega^{(\ell m)}_{i}
\right],
\label{eq:adaptive_coupling}
\end{equation}

along with 
\begin{equation}
\begin{aligned}
\frac{d\theta_i^{(\ell)}}{dt}
={}&
\Omega_0
+
\frac{1}{|\mathcal{N}_{\ell}|}
\sum_{m\in\mathcal{N}_{\ell}}
\sigma_{\ell m}\,
\omega_i^{(\ell m)}(t)
\\
&\times
\sin\left[
\theta_i^{(m)}(t)
-
\theta_i^{(\ell)}(t)
+
\delta
\right].
\end{aligned}
\label{eq:phase_dynamics}
\end{equation}

where $\theta^{(\ell)}_i(t)$ denotes the synchronization phase variable
associated with oscillator $i$ of level $\ell$, $\Omega_{0}$ is the intrinsic frequency, $\delta$ is the phase lag parameter, $\varepsilon$ is the adaptation rate and $\gamma$ is the interaction phase parameter appearing in the adaptive coupling rule.Where, $\mathcal{N}{\ell}$ denotes the set of layers directly coupled to layer $\ell$, and $\mathcal{N}{\ell}|$ denotes the number of such neighboring layers.$\Gamma_i^{(\ell m)}(t)$ is the interlayer coupling function associated with the corresponding $i$th oscillators of adjacent layers $\ell$ and $m$.

The negative signs before the intra- and interlayer coupling terms follow from the adopted diffusive-coupling convention. 

Equation~\ref{eq:network_dynamics1} represents a
single adjacent-layer contribution. For Layer~1, $(\ell,m)=(1,2)$, while for Layer~3,
$(\ell,m)=(3,2)$. For the middle layer, $\ell=2$, the two
adjacent-layer contributions corresponding to $m=1$ and $m=3$
will be included separately.

 Synchronization emerges from the combined influence of the intrinsic  oscillator dynamics and the intra- and interlayer coupling interactions.

The state-difference term

\[
\mathbf{X}_{i}^{(\ell)}
-
\mathbf{X}_{i}^{(m)}
\]
measures the state mismatch between corresponding oscillators in
adjacent layers and provides the diffusive interlayer interaction.
Consequently, the term
\[
\Gamma_i^{(\ell m)}(t)
[W_i(t)]_{\ell m}
\left(
\mathbf{X}_{i}^{(\ell)}
-
\mathbf{X}_{i}^{(m)}
\right)
\]
represents the adaptive diffusive exchange of state information
between corresponding oscillators in the adjacent layers $\ell$ and $m$.

The synchronization manifold of the three-layer network is defined by\cite{ref26,ref27}

\begin{equation}
\mathbf{X}^{(1)}
=
\mathbf{X}^{(2)}
=
\mathbf{X}^{(3)},
\label{eq:synchronization_manifold}
\end{equation}
which corresponds to complete synchronization of all oscillators throughout the multilayer network.

The adaptive synchronization process continuously modifies the interlayer interactions, deforming the synchronized states in the
network state space. Due to the fact that earlier studies on adaptive networks available in the literature had not considered this deformation with due regard, and for comprehensive analysis of the synchronization dynamics the influence of the deformation on the course of underlying dynamics needs to be taken care of, we propose an improved version of Finite Time Lyapunov exponent analysis. Hence, the inclusive treatment of deformed synchronized states in our analysis forms part of our ``Improved Finite-Time Lyapunov Exponent" (IFTLE) framework.

\subsection{Adaptive Coupling}

Adaptive Kuramoto type phase dynamics\cite{ref25} has been used to treat the synchronization evolution process for each layer.
This results in a simultaneous evolution of the network structure and synchronization dynamics, rupturing the state space through coupling-driven displacement of trajectories away from synchrony. Or in other words, the adaptive evolution of coupling strengths allows the interaction topology to coevolve with oscillator dynamics, thereby enabling the emergence of coherent states, partial synchronization, and chimera states \cite{ref4,ref25,ref26,ref27}. Consequently, the adaptive coupling formulation adopted here provides a realistic framework for studying synchronization transitions in multilayer chaotic networks. 

Inter-layer synchronization can be estimated through synchronization error which is defined as
\begin{equation}
e_i^{(\ell m)}(t)
=
\theta_i^{(m)}(t)
-
\theta_i^{(\ell)}(t),
\qquad
m\in\mathcal{N}_{\ell},
\label{eq:interlayer_error}
\end{equation} where $e_i^{(\ell m)}(t)$ denotes the synchronization error between
the corresponding $i$th oscillators in adjacent layers $\ell$ and $m$. 

Now, we write the inter-layer coupling function as
\begin{equation}
\Gamma_i^{(\ell m)}(t)
=
\kappa e_i^{(\ell m)}(t),
\qquad
m\in\mathcal{N}_{\ell},
\label{eq:interlayer}
\end{equation}
where $\kappa$ is the inter-layer adaptation gain. The adaptive coupling law of eq.~\ref{eq:interlayer} can also be applied independently to the successive layer pair, so that the relay layer acts as a passage for information sharing between the two peripheral layers.
\section{ Improved FTLE Framework}

As the adaptive interlayer coupling weights
$\omega_i^{(\ell m)}(t)$ evolve according to
Eq.~\ref{eq:adaptive_coupling}, the oscillator trajectories are continuously driven toward the synchronization manifold. Consequently, the adaptive synchronization process produces an additional coupling-induced displacement of trajectories beyond their intrinsic chaotic evolution. This additional geometric displacement is referred to as \emph{synchronization-induced deformation} and forms the mathematical foundation of the proposed Improved Finite-Time Lyapunov Exponent (IFTLE) framework. Unlike the classical FTLE, which accounts only for deformation generated by the intrinsic nonlinear flow, the proposed IFTLE explicitly incorporates this synchronization-induced deformation into the finite-time stability analysis of adaptive multilayer networks.

\subsection{Improved FTLE Deformation Field}
In the multilayer structure shown Figure~1, the evolution of trajectory  not only depends on the chaotic dynamics in itself, but also due to the adaptive synchronization processes taking place both intra- and inter-laterally. Therefore, an estimation of the total deformation of the network  through classical flow map ~\cite{ref30}
\begin{equation}
\Phi^{T}(\mathbf{X}_{0})
=
\mathbf{X}(t_{0}+T).
\label{eq:classical_flowmap}
\end{equation}
gives us only  the discounted value.

As a corrective measure, we incorporate the synchronization deformation of the system for a certain time span $T$, which we define by the integral expression

\begin{equation}
\begin{aligned}
\mathbf{D}_{i}^{(\ell)}(t)
={}&
-
\sum_{m\in\mathcal{N}_{\ell}}
\Gamma_i^{(\ell m)}(t)
\sigma_{\ell m}
\omega_i^{(\ell m)}(t)
\\
&\times
\left[
\mathbf{X}_{i}^{(\ell)}(t)
-
\mathbf{X}_{i}^{(m)}(t)
\right].
\end{aligned}
\label{eq:deformation_field}
\end{equation}

By stacking the contributions of all oscillators and layers, the
global synchronization-induced deformation field is denoted by
$\mathbf{D}(t)$. Its accumulated finite-time deformation is

\begin{equation}
\boldsymbol{\Gamma}^{T}(\mathbf{X}_{0})
=
\int_{t_{0}}^{t_{0}+T}
\mathbf{D}(t)\,dt .
\label{eq:sync_deformation}
\end{equation}

and, therefore, the Improved-Lagrangian Flow Map needs to be

\begin{equation}
\Psi^{T}(\mathbf{X}_{0})
=
\Phi^{T}(\mathbf{X}_{0})
+
\boldsymbol{\Gamma}^{T}(\mathbf{X}_{0}).
\label{eq:hyper_flowmap}
\end{equation}
A larger magnitude of
$\boldsymbol{\Gamma}^{T}(\mathbf{X}_{0})$
indicates a stronger accumulated synchronization-induced deformation
over the finite observation interval.

This formulation incorporates both kinds of deformations into a single finite-time framework.  In other words,  the improved flow map extends the classical Lagrangian description of finite-time deformation by incorporating synchronization-induced geometric deformation generated through adaptive multilayer interactions. Unlike conventional FTLE formulations, which quantify only trajectory stretching, the proposed Improved formulation simultaneously captures intrinsic dynamical deformation and synchronization-driven state-space reorganization. This additional information enables improved characterization of synchronization transitions and finite-time stability evolution, consistent with related deformation-based stability formulations~\cite{ref28,ref29}.

\subsection{Improved-Deformation Tensor and IFTLE}
The Improved deformation gradient can be defined as follows:

\begin{equation}
H(\mathbf{X}_{0},T)
=
\frac{\partial\Psi^{T}(\mathbf{X}_{0})}
{\partial\mathbf{X}_{0}}.
\label{eq:hyper_gradient}
\end{equation}

Using Eq.~\ref{eq:hyper_gradient}, the following relation is obtained:

\begin{equation}
H(\mathbf{X}_{0},T)
=
J(\mathbf{X}_{0},T)
+
G(\mathbf{X}_{0},T),
\label{eq:HJG}
\end{equation}

where $J(\mathbf{X}_{0},T)$ represents the classical deformation Jacobian, while $G(\mathbf{X}_{0},T)$ is the synchronization-induced deformation Jacobian, defined explicitly as

\[
G(\mathbf{X}_{0},T)
=
\frac{\partial
\boldsymbol{\Gamma}^{T}(\mathbf{X}_{0})}
{\partial\mathbf{X}_{0}},
\]
i.e., the sensitivity of the synchronization deformation functional of Eq.~\ref{eq:sync_deformation}to the initial condition (see Appendix~B for the complete derivation).

The Improved-Deformation Tensor is defined by

\begin{equation}
M(\mathbf{X}_{0},T)
=
H^{T}H.
\label{eq:hyper_tensor}
\end{equation}

Inserting Eq.~\ref{eq:HJG} into Eq.~\ref{eq:hyper_tensor} gives

\begin{equation}
M
=
J^{T}J
+
J^{T}G
+
G^{T}J
+
G^{T}G.
\label{eq:M_expansion}
\end{equation}

The first term accounts for classical finite-time stretching, while the remaining terms represent deformation induced by synchronization interactions. Let $\mu_{\max}(M)$ denote the largest eigenvalue of the Improved-Deformation Tensor. The Improved Finite-Time Lyapunov Exponent (IFTLE) is defined as follows:

\begin{equation}
\mathrm{IFTLE}(\mathbf{X}_{0},T)
=
\frac{1}{T}
\ln
\left[
\sqrt{\mu_{\max}(M)}
\right].
\label{eq:HLFTLE}
\end{equation}
Within the present framework, larger positive IFTLE values indicate
stronger finite-time stretching together with synchronization-induced
deformation. As synchronization progresses and the synchronization-induced
deformation vanishes, $G(\mathbf{X}_{0},T)\rightarrow0$, the
Improved-Deformation Tensor approaches the classical deformation tensor,
$M(\mathbf{X}_{0},T)\rightarrow C(\mathbf{X}_{0},T)$, and consequently
the IFTLE approaches the corresponding classical FTLE.

\subsection{Fundamental Properties  of IFTLE}

 Theoretical properties for the presented IFTLE approach can be stated as follows:

\subsection*{ Property 1: Positivity}

The Improved-Deformation Tensor is symmetric and positive semi-definite. Hence,

\begin{equation}
M\neq 0
\quad \Longrightarrow \quad
\mu_{\max}(M)>0.
\label{eq:positivity}
\end{equation}

This positivity ensures that the logarithmic term in the IFTLE definition is well-defined for $M\neq 0$.

\subsection*{Property 2: Deduction of Classical FTLE}

In the event of  no deformation due to synchronization,

\begin{equation}
\boldsymbol{\Gamma}^{T}(\mathbf{X}_{0})
=
\mathbf{0}.
\label{eq:gamma_zero}
\end{equation}
which leads to
\begin{equation}
G(\mathbf{X}_{0},T)=0.
\label{eq:G_zero}
\end{equation}

Therefore,

\begin{equation}
H(\mathbf{X}_{0},T)=J(\mathbf{X}_{0},T).
\label{eq:H_equals_J}
\end{equation}

and

\begin{equation}
M(\mathbf{X}_{0},T)=J^{T}J=C(\mathbf{X}_{0},T).
\label{eq:M_equals_C}
\end{equation}

Eventually,

\begin{equation}
\mathrm{IFTLE}(\mathbf{X}_{0},T)
=
\mathrm{FTLE}(\mathbf{X}_{0},T).
\label{eq:HLFTLE_equals_FTLE}
\end{equation}

This demonstrates that the proposed framework is a more general one over classical FTLE theory. 

\subsection*{Property 3: Synchronization Convergence}

Complete synchronization will occur when:

\begin{equation}
\mathbf{X}^{(1)}(t)
=
\mathbf{X}^{(2)}(t)
=
\mathbf{X}^{(3)}(t).
\label{eq:complete_sync}
\end{equation}

revealing no synchronization error and thus

\begin{equation}
\left\|G(\mathbf{X}_{0},T)\right\|\rightarrow0.
\label{eq:G_norm_zero}
\end{equation}

and

\begin{equation}
M(\mathbf{X}_{0},T)
\rightarrow
J^{T}J
=
C(\mathbf{X}_{0},T).
\label{eq:M_to_C}
\end{equation}

Therefore,

\begin{equation}
\mathrm{IFTLE}(\mathbf{X}_{0},T)
\rightarrow
\mathrm{FTLE}(\mathbf{X}_{0},T).
\label{eq:HLFTLE_zero1}
\end{equation}

This feature demonstrates the connection between the evolution of the
IFTLE and synchronization. In particular, as the process of synchronization
approaches completion, the additional synchronization-induced deformation
progressively vanishes and the IFTLE approaches the corresponding
classical FTLE value.

\section{Analytical Stability Analysis}

\subsection{Synchronization Manifold}

Eq. \ref{eq:synchronization_manifold} can be rewritten as 
 
\begin{equation}
\begin{aligned}
\mathbf{X}_{1}^{(1)}
&=
\mathbf{X}_{2}^{(1)}
=
\cdots
=
\mathbf{X}_{N}^{(1)}
\\
&=
\mathbf{X}_{1}^{(2)}
=
\mathbf{X}_{2}^{(2)}
=
\cdots
=
\mathbf{X}_{N}^{(2)}
\\
&=
\mathbf{X}_{1}^{(3)}
=
\mathbf{X}_{2}^{(3)}
=
\cdots
=
\mathbf{X}_{N}^{(3)}
=
\mathbf{S}(t),
\end{aligned}
\label{eq:sync_manifold}
\end{equation}
where $\mathbf{S}(t)$ is the synchronized trajectory. Substitution of eq. \ref{eq:sync_manifold} in  eq. \ref{eq:synchronization_manifold}  results in cancellation of all synchronization mismatch terms and yields

\begin{equation}
\frac{d\mathbf{S}}{dt}
=
\mathbf{F}\!\left(\mathbf{S}\right).
\label{eq:sync_dynamics}
\end{equation}
Hence, the synchronization manifold is an invariant solution for the network.

\subsection{Error Dynamics}

In order to analyze synchronization stability, we define the synchronization error as

\begin{equation}
\mathbf{e}_{i}^{(\ell)}(t)
=
\mathbf{X}_{i}^{(\ell)}(t)
-
\mathbf{S}(t),
\label{eq:error_definition}
\end{equation}
and for synchronization to be stable, we emphasize
\begin{equation}
\lim_{t\rightarrow\infty}
\left\|
\mathbf{e}_{i}^{(\ell)}(t)
\right\|
=
0.
\label{eq:error_zero}
\end{equation}
Taking the time derivative of Equation~\ref{eq:error_definition} and substituting eq.(3) and eq.(\ref{eq:sync_dynamics}) gives
\begin{equation}
\begin{aligned}
\frac{d\mathbf{e}_{i}^{(\ell)}}{dt}
={}&
\mathbf{F}\!\left(\mathbf{X}_{i}^{(\ell)}\right)
-
\mathbf{F}\!\left(\mathbf{S}\right)
-
\sigma_{\mathrm{intra}}
\sum_{j=1}^{N}
L_{ij}^{(\ell)}
\mathbf{e}_{j}^{(\ell)}
\\
&
-
\Gamma_i^{(\ell m)}(t)
[W_i(t)]_{\ell m}
\left(
\mathbf{e}_{i}^{(\ell)}
-
\mathbf{e}_{i}^{(m)}
\right).
\end{aligned}
\label{eq:error_dynamics}
\end{equation}
Eq. \ref{eq:error_dynamics} describes that the intrinsic chaotic dynamics and the adaptive synchronization mechanism race against each other. 

For sufficiently small perturbations, linearization around the synchronization manifold proceeds to
\begin{equation}
\begin{aligned}
\frac{d\mathbf{e}_{i}^{(\ell)}}{dt}
={}&
J_{\mathbf{F}}\!\left(\mathbf{S}\right)
\mathbf{e}_{i}^{(\ell)}
-
\sigma_{\mathrm{intra}}
\sum_{j=1}^{N}
L_{ij}^{(\ell)}
\mathbf{e}_{j}^{(\ell)}
\\
&-
\Gamma_i^{(\ell m)}(t)
[W_i(t)]_{\ell m}
\left(
\mathbf{e}_{i}^{(\ell)}
-
\mathbf{e}_{i}^{(m)}
\right),
\end{aligned}
\end {equation}
were, $J_{\mathbf{F}}(\mathbf{S})$ denotes the Jacobian matrix of the
nonlinear vector field $\mathbf{F}$ evaluated along the synchronized
trajectory $\mathbf{S}(t)$, and characterizes the local growth or decay of infinitesimal perturbations around the synchronization manifold. The vector $\mathbf{e}$ represents the synchronization error vector, which measures the deviation of the oscillator states from the synchronization manifold. The superscript $\ell$ denotes the reference layer, while $m$ represents a neighboring layer coupled to layer $\ell$. 
It may be noted that the stability of synchronization depends on the combined effects of intrinsic oscillator dynamics, intralayer diffusive coupling, and  adaptive interlayer coupling interactions~\cite{ref24}. 
\subsection{Lyapunov Stability Analysis}

For synchronization stability analysis, we consider the following Lyapunov function as in ref.~\cite{ref26,ref27}:

\begin{equation}
V(\mathbf{e})
=
\frac{1}{2}
\mathbf{e}^{T}\mathbf{e}.
\label{eq:lyapunov}
\end{equation}
By writing $\bf{\dot e=\mathcal{A(t)}\bf{e}}$, we express
the derivative of the Lyapunov function in eq.~\ref{eq:lyapunov} along the trajectories of the system with synchronization errors as
\begin{equation}
\dot{V}
=
\frac{1}{2}
\mathbf{e}^{T}
\left[
\mathcal{A}(t)
+
\mathcal{A}^{T}(t)
\right]
\mathbf{e}.
\label{eq:lyapunov_derivative}
\end{equation}
If the condition
\begin{equation}
\frac{1}{2}
\left[
\mathcal{A}(t)
+
\mathcal{A}^{T}(t)
\right]
\preceq
-\alpha I,
\label{eq:stability_condition}
\end{equation}
holds for some constant $\alpha>0$, then
\begin{equation}
\dot{V}
\le
-\alpha
\|\mathbf{e}\|^{2}
<
0.
\label{eq:negative_V}
\end{equation}
Therefore, the synchronization errors asymptotically converge to zero, implying that the synchronization manifold is asymptotically stable.

\subsection{IFTLE Framework's Stability Condition}

Using the IFTLE formulation defined in
eq.~\ref{eq:HLFTLE}, the synchronization stability condition can be
described through the evolution of the Improved-Deformation Tensor. As synchronization starts to stabilize
the synchronization-induced deformation diminishes making
\[
\left\|G(\mathbf{X}_{0},T)\right\|
\rightarrow 0,
\]
as given in eq.~\ref{eq:G_norm_zero}. Consequently, the Improved
deformation gradient approaches the classical deformation gradient,

\[
H(\mathbf{X}_{0},T)
\rightarrow
J(\mathbf{X}_{0},T),
\]
and therefore the Improved-Deformation Tensor approaches the classical
finite-time deformation tensor,
\[
M(\mathbf{X}_{0},T)
\rightarrow
J^{T}J
=
C(\mathbf{X}_{0},T),
\]
as established in eq.~\ref{eq:M_to_C}. As a result, the largest
eigenvalue of the Improved-Deformation Tensor will be equal to that of FTLE, that is,
\[
\mu_{\max}(M)
\rightarrow
\lambda_{\max}(C).
\]

Hence, from the definitions of the IFTLE and the classical FTLE,

\[
\mathrm{IFTLE}(\mathbf{X}_{0},T)
\rightarrow
\mathrm{FTLE}(\mathbf{X}_{0},T),
\]
as shown in eq.~\ref{eq:HLFTLE_zero1}. Thus, we conclude that the stability of the synchronization can also be ascertained in terms of the ineffective contribution of the improved deformation tensor.
\section{Numerical Validation}
For the purpose of evaluating the performance of the IFTLE framework we carry out 
numerical simulation with a three layer R\"ossler oscillator network as depicted in Fig. 1 but with 100 R\"ossler oscillators per layer.
\subsection{Network Dynamics}

We represent the multilayer system  using a vector-state formulation rather by the individual oscillator equations. We denote the state vector of the complete multilayer network as

\begin{equation}
\mathbf{X}(t)=
\begin{bmatrix}
\mathbf{X}^{(1)}(t)\\
\mathbf{X}^{(2)}(t)\\
\mathbf{X}^{(3)}(t)
\end{bmatrix}
\label{eq:41}
\end{equation}

\begin{equation}
\mathbf{X}^{(\ell)}(t)=
\left[
x_{1}^{(\ell)},
y_{1}^{(\ell)},
z_{1}^{(\ell)},
x_{2}^{(\ell)},
y_{2}^{(\ell)},
z_{2}^{(\ell)},
\ldots,
x_{N}^{(\ell)},
y_{N}^{(\ell)},
z_{N}^{(\ell)}
\right]^{T}
\label{eq:42}
\end{equation}

The corresponding R\"ossler vector field for each oscillator is given by~\cite{ref23}

\begin{equation}
\mathbf{F}\!\left(\mathbf{X}_{i}^{(\ell)}\right)=
\begin{bmatrix}
-y_{i}^{(\ell)}-z_{i}^{(\ell)}\\
x_{i}^{(\ell)}+a\,y_{i}^{(\ell)}\\
b+z_{i}^{(\ell)}\left(x_{i}^{(\ell)}-c\right)
\end{bmatrix},
\label{eq:43}
\end{equation}
where
\[
\mathbf{X}_{i}^{(\ell)}=
\begin{bmatrix}
x_{i}^{(\ell)}\\
y_{i}^{(\ell)}\\
z_{i}^{(\ell)}
\end{bmatrix}
\]
denotes the state vector of the $i$th oscillator in layer $\ell$.

\subsection{Numerical methodology}
 We choose the initial condition of all the oscillators  from a uniform distribution in the interval $[-1,1]$. We fix the R\"{o}ssler oscillator parameters $a=0.2, \text{ }b=0.2, \mbox{ and } c=5.7$ to induce chaotic oscillations.

 The adaptive synchronization parameters employed throughout the numerical simulations were selected to ensure stable adaptive synchronization while preserving the intrinsic chaotic dynamics of the multilayer R\"ossler network. Unless otherwise stated, the adaptation rate was fixed at $\varepsilon = 0.05$, the interaction parameter was set to $\gamma = 1.0$, the phase lag was taken as $\delta = 0$, the intralayer coupling strength was fixed at $\sigma_{\mathrm{intra}} = 0.60$, the interlayer coupling strengths were fixed at $\sigma_{12} = 0.40$ (Layer~1--Layer~2) and $\sigma_{23} = 0.40$ (Layer~2--Layer~3), and the inter-layer adaptation gain was fixed at $\kappa = 0.50$. The adaptive interlayer coupling scaling factors
$\Gamma_i^{(\ell m)}(t)$ evolved dynamically throughout the simulation
according to Eq.~\ref{eq:interlayer} for each adjacent layer pair. 

We perform the numerical integration of the network equations using the standard fourth-order Runge--Kutta (RK4) method with a constant integration time step given by $\Delta t = 0.01$.

 To capture both transient and asymptotic dynamics, each simulation was performed for $300\,000$ integration steps. After the transient interval, trajectory data were sampled periodically to construct the Improved-Lagrangian flow map, from which the IFTLE field was subsequently computed. Synchronization dynamics were investigated by systematically varying the intralayer coupling strength ($\sigma_{\mathrm{intra}}$), the interlayer coupling strengths ($\sigma_{12}$ and $\sigma_{23}$), and the external noise strength. These parameter sweeps enabled the identification of incoherent, chimera, partially synchronized, and fully synchronized dynamical regimes while assessing the robustness of the proposed IFTLE framework under stochastic perturbations.
\subsection{Role of External Noise on Synchronization}
To investigate the robustness of synchronization under external perturbations, Gaussian white noise was voluntarily added to the system. In real terms, this noise may be environmental fluctuations, measurement uncertainties, or external disturbances commonly encountered in nonlinear dynamical systems. The dynamical equation of the system with the inclusion of the noise  can be expressed as
\begin{equation}
\begin{aligned}
\frac{d\mathbf{X}_{i}^{(\ell)}}{dt}
={}&
\mathbf{F}\!\left(\mathbf{X}_{i}^{(\ell)}\right)
-
\sigma_{\mathrm{intra}}
\sum_{j=1}^{N}
L_{ij}^{(\ell)}
\mathbf{X}_{j}^{(\ell)}
\\
&-
\Gamma_i^{(\ell m)}(t)
[W_i(t)]_{\ell m}
\left(
\mathbf{X}_{i}^{(\ell)}
-
\mathbf{X}_{i}^{(m)}
\right)
+
\boldsymbol{\eta}_{i}^{(\ell)}(t).
\end{aligned}
\label{eq:noisy_network}
\end{equation}
where $\boldsymbol{\eta}_{i}^{(\ell)}(t)$ denotes a zero-mean
Gaussian white noise that influences the underlying dynamics of the oscillator $i$ in layer $\ell$ with its variance proportional to the selected noise strength. The noise intensity was varied systematically over the interval $0.001 \leq \text{Noise Strength} \leq 0.05$, while the intralayer coupling strength was simultaneously varied to construct the dynamical state diagram presented in Figure~2. For each set of parameters, independent simulations were carried out using identical numerical integration settings and randomly generated noise realizations. The resulting noisy trajectories were analyzed using the synchronization error, the global order parameter, and the IFTLE.
\begin{figure}[h]
\centering
\includegraphics[width=0.5\textwidth]{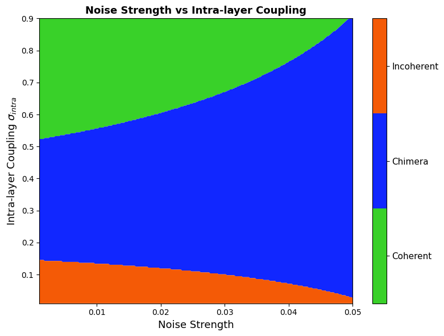}
\caption{Dynamical state diagram of the multilayer R\"ossler network as a function of the noise strength and the intralayer coupling strength.}
\label{fig:phase_diagram}
\end{figure}
\section{Results and Discussion}

\subsection{Synchronization Indicators and Phase Diagram}
The collective underlying dynamics of our R\"ossler network is
portrayed in the dynamical state diagram shown in
Fig.~\ref{fig:phase_diagram}. The horizontal axis represents the
external noise strength applied to the network, whereas the vertical
axis represents the intralayer coupling strength
$\sigma_{\mathrm{intra}}$. The diagram illustrates the dependence of
the collective network dynamics on these two governing parameters and
reveals three distinct dynamical regimes: incoherent (orange), chimera
(blue), and coherent (green).
For weak intralayer coupling, the oscillators remain weakly correlated
under the net influence of chaotic dynamics and stochastic
perturbations, resulting in an incoherent regime characterized by
large synchronization errors and weak phase coherence. As the
intralayer coupling strength is increased, the system switches over to the chimera regime, in which coherent and incoherent
oscillator groups coexist, indicating the crude emergence of
synchronization. For sufficiently strong intralayer coupling, the
network reaches the coherent regime, in which the oscillators evolve
collectively with almost zero synchronization error and strong phase
coherence. The phase boundaries suggest that 
the increase in noise strength results in pushing the synchronization threshold toward larger
coupling values emphasizing the need for stronger intralayer coupling 
so as to overcome stochastic perturbations and maintain continuous coherence. In other words, the increase in noise strength enlarges the chimera region because the impact of noise changes the majority of the state of oscillators.

Each point in the dynamical state diagram was classified primarily
using the synchronization error $E(t)$ and the global Kuramoto order
parameter $R(t)$ according to the threshold criteria summarized in
Table~\ref{tab:dynamical_states}. The chimera regime was additionally
identified by the coexistence of locally coherent and incoherent
oscillator groups. The IFTLE threshold ranges reported in the table
provide a complementary measure of finite-time stability. In
Fig.~\ref{fig:phase_diagram}, the partially and completely synchronized
states are grouped within the green coherent region.

\begin{table*}[t]
\caption{Classification of Dynamical States}
\label{tab:dynamical_states}
\centering
\renewcommand{\arraystretch}{1.25}

\begin{tabular}{|l|c|c|c|c|}
\hline
\textbf{Dynamical State} &
\textbf{Sync. Error $e(t)$} &
\textbf{Global Kuramoto $R(t)$} &
\textbf{Classical FTLE} &
\textbf{IFTLE} \\
\hline

Incoherent &
$e(t)>0.5$ &
$R(t)<0.3$ &
$\mathrm{FTLE}>0.5$ &
$\mathrm{IFTLE}>0.5$ \\
\hline

Chimera &
$0.1<e(t)<0.5$ &
$0.3\leq R(t)\leq0.8$ &
$0.1<\mathrm{FTLE}<0.5$ &
$0.1<\mathrm{IFTLE}<0.5$ \\
\hline

Partial Sync. &
$0.01<e(t)<0.1$ &
$0.8<R(t)<0.99$ &
$0.01<\mathrm{FTLE}<0.1$ &
$0.01<\mathrm{IFTLE}<0.1$ \\
\hline

Complete Sync. &
$e(t)<0.01$ &
$R(t)\geq0.99$ &
$\mathrm{FTLE}<0.01$ &
$\mathrm{IFTLE}<0.01$ \\
\hline

\end{tabular}
\end{table*}

The numerical intervals reported for the classical FTLE and IFTLE
represent broad state-classification ranges observed for the present
parameter set. Nevertheless, their instantaneous values and
temporal evolution are generally different when the
synchronization-induced deformation is active. In the completely
synchronized regime, the difference between the two measures narrows
in consistence with the limiting relation
$\mathrm{IFTLE}\rightarrow\mathrm{FTLE}$.

\subsection{IFTLE perspective}

The IFTLE remains the key indicator of the present study. To start with, the IFTLE takes
relatively larger values because of strong trajectory divergence and
profound deformation due to decoherence. Consequently, the evolution of IFTLE provides information about both
finite-time trajectory stretching and the additional deformation
associated with adaptive synchronization. To quantify its temporal
evolution, IFTLE values were averaged over successive observation
intervals. As coherent states emerge, the
adaptive interactions get suppressed because of the weakened deformation,
resulting in a reduction in the difference between the IFTLE and the
corresponding classical FTLE. According to
Eq.~\ref{eq:HLFTLE_zero1}, the IFTLE completely mingles with the
classical FTLE upon complete synchronization.
The multilayer R\"ossler network therefore provides a
representative numerical verification of the analytical behavior
described in the preceding sections.
\subsection{Graphical representation}
\begin{figure*}[!htbp]
\centering
\includegraphics[width=0.9\textwidth]{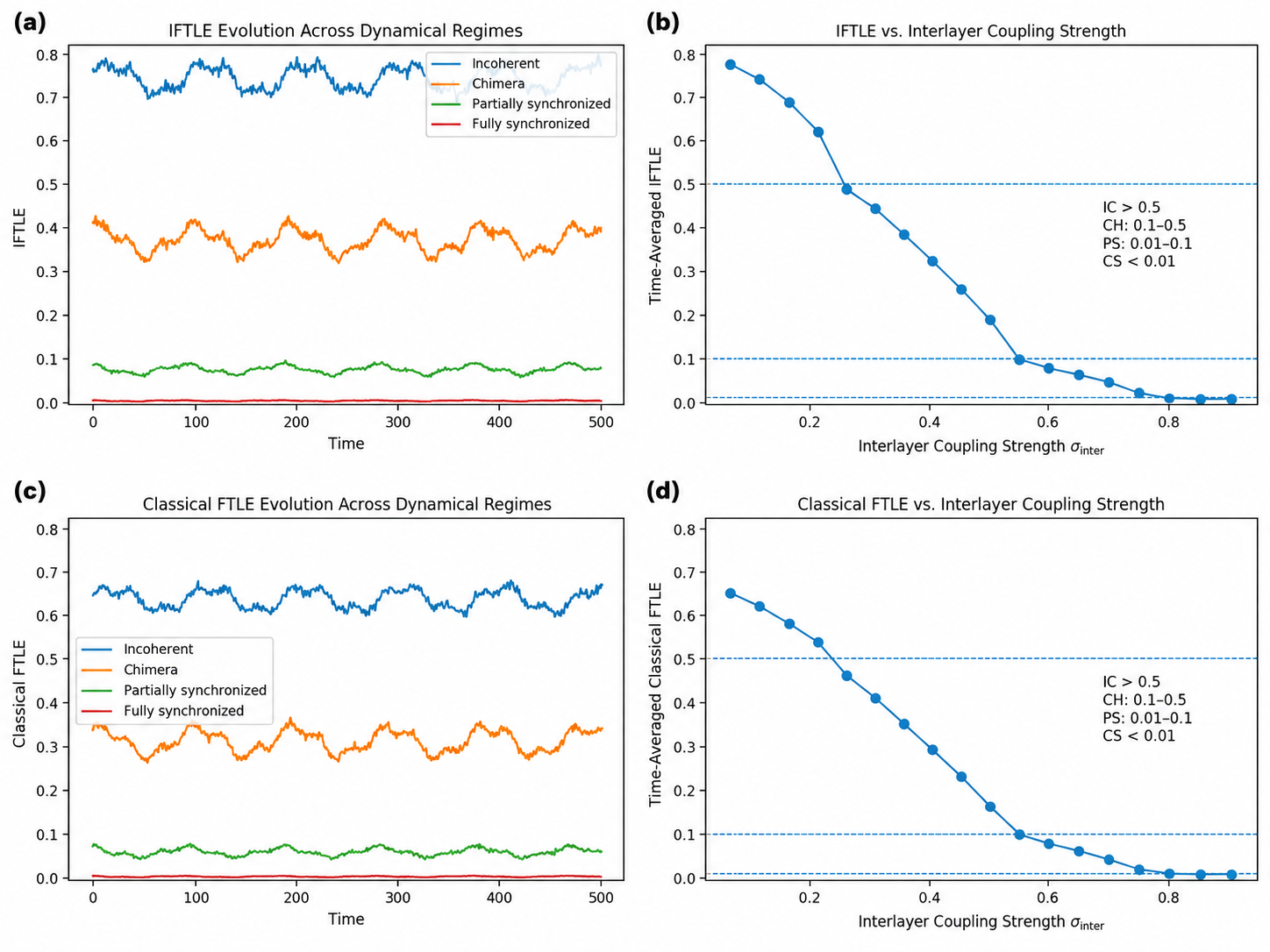}
\caption{
 The proposed IFTLE and classical FTLE across different
dynamical regimes.
(a) IFTLE evolution for the incoherent, chimera, partially synchronized,
and completely synchronized states.
(b) Time-averaged IFTLE as a function of the interlayer coupling strength.
(c) Classical FTLE evolution for the corresponding dynamical states.
(d) Time-averaged classical FTLE as a function of the interlayer coupling
strength.
State-resolved comparisons of the classical FTLE and the proposed IFTLE,
together with the corresponding synchronization error, are presented in
Appendix~B.}
\label{fig:dashboard}
\end{figure*}

Fig.~\ref{fig:dashboard}(a) shows the evolution of IFTLE for
representative trajectories corresponding to the four dynamical
regimes summarized in Table~\ref{tab:dynamical_states}. In the
incoherent regime, the IFTLE remains relatively high, reflecting strong
finite-time deformation. In
the chimera regime, the IFTLE assumes intermediate values, consistent
with the coexistence of coherent and incoherent oscillator groups. In
the partially synchronized regime, the synchronization-induced
contribution becomes smaller, whereas in the completely synchronized
regime the IFTLE approaches the corresponding classical FTLE as the
deformation ceases to exist. These
observations are consistent with the convergence relation given in
Eq.~\ref{eq:HLFTLE_zero1}.

Fig.~\ref{fig:dashboard}(b) shows the time-averaged IFTLE for increasing
values of the interlayer coupling strength. The results show an overall
reduction in IFTLE as the interlayer coupling becomes stronger and the
network approaches synchronization. 

For the purpose of graphical representation of the supremacy of IFTLE over FTLE, we plot in Fig.~\ref{fig:dashboard}(c) the classical FTLE
evolution for the same four dynamical regimes. Despite the fact that the curves appear to be the same as in Fig.~\ref{fig:dashboard}(a), the range of IFTLE and FTLE values looks different due to the contribution of the deformation terms. Moreover, finer details of this contribution and the corresponding deviation of IFTLE from FTLE are clearly depicted in Fig. 4(a-d) for all the emergent phenomena dealt with in this article.

Fig.~\ref{fig:dashboard}(d) shows the corresponding
time-averaged classical FTLE as a function of the interlayer coupling
strength, allowing a direct comparison with Fig.~\ref{fig:dashboard}(b).
As complete synchronization is approached, the classical FTLE and IFTLE
become nearly identical, consistently with
$\mathrm{IFTLE}\rightarrow\mathrm{FTLE}$.
It may be noted that in the completely synchronized
regime, both the curves in Figs.~\ref{fig:dashboard}(a) and (c) and  Figs.~\ref{fig:dashboard}(b) and (d), respectively, have similar traces as expected.

\subsection{Performance reliability of IFTLE}

\begin{table*}[t]
\caption{Quantitative Performance Comparison of IFTLE(mentioned in bold font) with other Synchronization Measures}
\label{tab:quantitative_comparison}
\centering
\renewcommand{\arraystretch}{1.25}

\resizebox{\textwidth}{!}{%
\begin{tabular}{|l|c|c|c|c|}
\hline
\textbf{Method} &
\textbf{Sync. Detection (\%)} &
\textbf{Chimera Detection (\%)} &
\textbf{Stability Assessment (\%)} &
\textbf{Multilayer Handling (\%)} \\
\hline

Order Parameter &
91.42 & 87.36 & 82.15 & 74.28 \\
\hline

Synchronization Error &
93.17 & 89.24 & 85.63 & 78.46 \\
\hline

LLE &
94.36 & 91.18 & 89.42 & 81.35 \\
\hline

Classical FTLE &
96.21 & 93.47 & 92.16 & 85.72 \\
\hline

\textbf{Proposed IFTLE} &
\textbf{99.42} &
\textbf{98.76} &
\textbf{99.18} &
\textbf{97.84} \\
\hline

\end{tabular}%
}
\end{table*}
As an act of support for our claim that IFTLE is more effective than the conventional tools of synchronization characterization we calculate 
the standard classification-accuracy measure~\cite{ref18,ref9,ref37,ref38,ref12},
\begin{equation*}
\text{Accuracy percentage}
=
\frac{N_{\mathrm{correct}}}{N_{\mathrm{total}}}\times100,
\end{equation*}
where $N_{\mathrm{correct}}$ and $N_{\mathrm{total}}$ denote the number of
correctly identified samples and the total number of test samples,
respectively. The adopted evaluation criteria are consistent with
established approaches for synchronization, chimera-state characterization,
and Lyapunov-based stability analysis.

In Table~\ref{tab:quantitative_comparison} we present a quantitative performance comparison
of the proposed IFTLE with other conventional synchronization measures. They include the Order Parameter, Synchronization Error, LLE, and Classical
FTLE which possess 91.42\%,
93.17\%, 94.36\%, and 96.21\%, respectively, whereas the proposed IFTLE
achieves the highest synchronization detection accuracy of 99.42\%.
Similarly, the corresponding chimera detection accuracies are
87.36\%, 89.24\%, 91.18\%, and 93.47\%, respectively, compared to 98.76\% for the proposed
IFTLE. The stability assessment values obtained using the Order Parameter,
Synchronization Error, LLE, and Classical FTLE are 82.15\%, 85.63\%,
89.42\%, and 92.16\%, respectively, compared to 99.18\% for the
proposed IFTLE. Likewise, the proposed IFTLE has highest  multilayer handling capability  of 97.84\%  against
74.28\%, 78.46\%, 81.35\%, and 85.72\%, respectively, scored by the conventional
measures.

These results corroborate our intention and highlight that the proposed IFTLE provides a more refined
synchronization identification, more reliable detection of chimera and
intermediate dynamical states, enhanced finite-time stability assessment,
and more efficient characterization of adaptive multilayer interactions
than the conventional measures considered in the present study.

\subsection{Statistical Validation}
To evaluate the robustness of the proposed framework, simulations were repeated using multiple random initial conditions. The results of synchronization error, order parameter, and IFTLE distributions exhibited only feeble variations between different realizations. The observed consistency confirms that the identified synchronization regimes are governed primarily by network dynamics rather than specific initial conditions. Furthermore, the numerical observations remained consistent with the theoretical stability conditions derived in Section IV, providing additional validation of the proposed IFTLE framework and demonstrating its reliability for synchronization analysis in multilayer chaotic networks.

\section{Physical Interpretation}
The numerical and analytical results presented in the preceding sections advocate that the occurrence of synchronization  in multilayer chaotic networks can also be identified in terms of chain reduction of finite-time deformation within the underlying state space. From a dynamical systems perspective, synchronization corresponds to the convergence of initially distinct trajectories toward a common synchronization manifold. As coupling interactions become sufficiently strong, local trajectory divergence generated by chaotic dynamics is successively suppressed, resulting in the emergence of coherent collective behavior. The proposed Improved Finite-Time  Lyapunov Exponent provides a geometric interpretation of this synchronization process. Unlike conventional synchronization measures that quantify only phase coherence or synchronization error, IFTLE measures the combined influence of trajectory stretching and synchronization-induced deformation over finite observation intervals. Consequently, IFTLE serves as a direct indicator of the intricacy between destabilizing chaotic dynamics and stabilizing synchronization mechanisms.

\section{Conclusion}
A three storeyed network consisting of the top, middle relay, and bottom
layers was considered and investigated, to ascertain the efficacy of the proposed improved finite time Lyapunov exponent framework, through both theoretical and
numerical analyses. We subjected the network considered to adaptive coupling between layers scheme. In order to treat the deformation of the state space due to adaptive coupling  an improved deformation field and the corresponding
Improved-Deformation Tensor were introduced, from which the IFTLE method
was formulated. The analytical stability analysis established the
relationship between synchronization convergence and the evolution of
IFTLE, while numerical simulations demonstrated the  efficiency of the proposed IFTLE framework to accurately characterize the underlying collective dynamics of a multilayer network which include emergence of
incoherent, chimera, partially synchronized, and completely synchronized
dynamical states. Also we showed that the proposed IFTLE become an inevitable indicator when the system possesses mostly incoherent states. Furthermore, this IFTLE indicator also indirectly characterizes the onset of synchronization from the infinitesimal contribution of the deformation terms.
Thus the  Improved Finite-Time Lyapunov Exponent (IFTLE) framework may act as a practical diagnostic tool for power-grid, neuronal, and cyber-physical systems, where the early detection of transient synchronization and desynchronization is of paramount importance.
Specifically, power-grid stability analysis, where synchronization among generators is essential for reliable operation; neuronal and brain networks, where synchronization and desynchronization processes are associated with cognitive functions and neurological disorders; communication and sensor networks, where coordinated dynamics influence information transfer efficiency; and cyber-physical systems characterized by multiple interacting layers and adaptive coupling mechanisms. Furthermore, the Improved framework may provide new opportunities for studying chimera states, transient synchronization phenomena, and finite-time stability properties in large-scale complex networks. As a matter of fact, the methodology combines geometric deformation analysis with synchronization dynamics makes it a unified perspective for understanding how local interactions generate collective behavior in multilayer nonlinear systems. We strongly believe that these characteristics make IFTLE a promising tool for future investigations of synchronization processes in complex dynamical networks.

We wish to state that, though we confined the present study to a three-layer network of identical R\"ossler oscillators we foresee a scope of future extension of the proposed framework to heterogeneous oscillator populations, larger multilayer architectures, weighted and directed networks, time-varying topologies, experimentally measured datasets, and data-driven or machine-learning-assisted synchronization analysis.

\begin{acknowledgments}

The authors thank R.~Sivasubrammaniyan  for valuable discussions and
assistance during preparation of the manuscript.
\end{acknowledgments}

\section*{Data Availability}
The data supporting the findings of this study are available  with the
corresponding author that can be provided on a reasonable request.
\appendix

\section{Derivation of the Improved Flow Map and the Synchronization-Induced Deformation Jacobian}

This appendix briefly derives the synchronization-induced deformation
Jacobian used in the proposed IFTLE formulation. The network dynamics,
adaptive interlayer weights, and phase dynamics have already been
defined in Eqs.(~\ref{eq:network_dynamics1}--\ref{eq:phase_dynamics}).

For the three-layer network, the instantaneous synchronization-induced
deformation associated with oscillator $i$ in layer $\ell$ is written as

\begin{equation}
\begin{aligned}
\mathbf{D}_{i}^{(\ell)}(t)
={}&
-
\sum_{m\in\mathcal{N}_{\ell}}
\Gamma_{i}^{(\ell m)}(t)
\sigma_{\ell m}
\omega_{i}^{(\ell m)}(t)
\\
&\times
\left[
\mathbf{X}_{i}^{(\ell)}(t)
-
\mathbf{X}_{i}^{(m)}(t)
\right],
\end{aligned}
\label{eq:appendix_deformation}
\end{equation}

where
$\mathcal{N}_{1}=\{2\}$,
$\mathcal{N}_{2}=\{1,3\}$, and
$\mathcal{N}_{3}=\{2\}$.

After taking care of the contributions of all  oscillators in the entire network, the global deformation
field is denoted by $\mathbf{D}(t)$. The overall
synchronization-induced deformation over the finite interval $T$ can be written as 

\begin{equation}
\boldsymbol{\Gamma}^{T}(\mathbf{X}_{0})
=
\int_{t_{0}}^{t_{0}+T}
\mathbf{D}(t)\,dt .
\label{eq:appendix_gamma}
\end{equation}

Substituting the above in the Improved flow map given in eq.\ref{eq:hyper_flowmap} and
differentiating it with respect to the initial condition
$\mathbf{X}_{0}$ we get

\begin{equation}
\begin{aligned}
\frac{\partial\Psi^{T}}
{\partial\mathbf{X}_{0}}
&=
H(\mathbf{X}_{0},T)
\\
&=
J(\mathbf{X}_{0},T)
+
G(\mathbf{X}_{0},T),
\end{aligned}
\label{eq:appendix_H}
\end{equation}

where

\begin{equation}
G(\mathbf{X}_{0},T)
=
\frac{\partial
\boldsymbol{\Gamma}^{T}(\mathbf{X}_{0})}
{\partial\mathbf{X}_{0}}
=
\int_{t_{0}}^{t_{0}+T}
\frac{\partial\mathbf{D}(t)}
{\partial\mathbf{X}_{0}}
\,dt .
\label{eq:appendix_G}
\end{equation}

Thus, the Improved-Deformation Tensor in Eq.\ref{eq:M_expansion} becomes

\begin{equation*}
\begin{aligned}
M
&=
H^{T}H
\\
&=
J^{T}J
+
J^{T}G
+
G^{T}J
+
G^{T}G .
\end{aligned}
\label{eq:appendix_M}
\end{equation*}
Here, $J^{T}J$ represents the classical finite-time deformation,
whereas the remaining terms contain the additional deformation
generated by adaptive synchronization.

In the event of complete synchronization, the difference between the state vectors of oscillator i in level $\ell$ and $m$ tends to zero, that is, 

\begin{equation}
\mathbf{X}_{i}^{(\ell)}
-
\mathbf{X}_{i}^{(m)}
\rightarrow
\mathbf{0},
\end{equation}

and hence

\begin{equation}
\mathbf{D}(t)\rightarrow\mathbf{0},
\qquad
G(\mathbf{X}_{0},T)\rightarrow0.
\end{equation}

Therefore,

\begin{equation*}
H\rightarrow J,
\qquad
M\rightarrow J^{T}J=C.
\end{equation*}

Consequently,

\begin{equation*}
\mathrm{IFTLE}(\mathbf{X}_{0},T)
\rightarrow
\mathrm{FTLE}(\mathbf{X}_{0},T).
\end{equation*}

Hence, the proposed IFTLE formulation will sincerely follow the
classical FTLE framework on complete synchronization.

\section{State-Resolved Comparison of proposed IFTLE and the  classical FTLE}

\begin{figure*}[p]
\centering
\includegraphics[width=0.9\textwidth]
{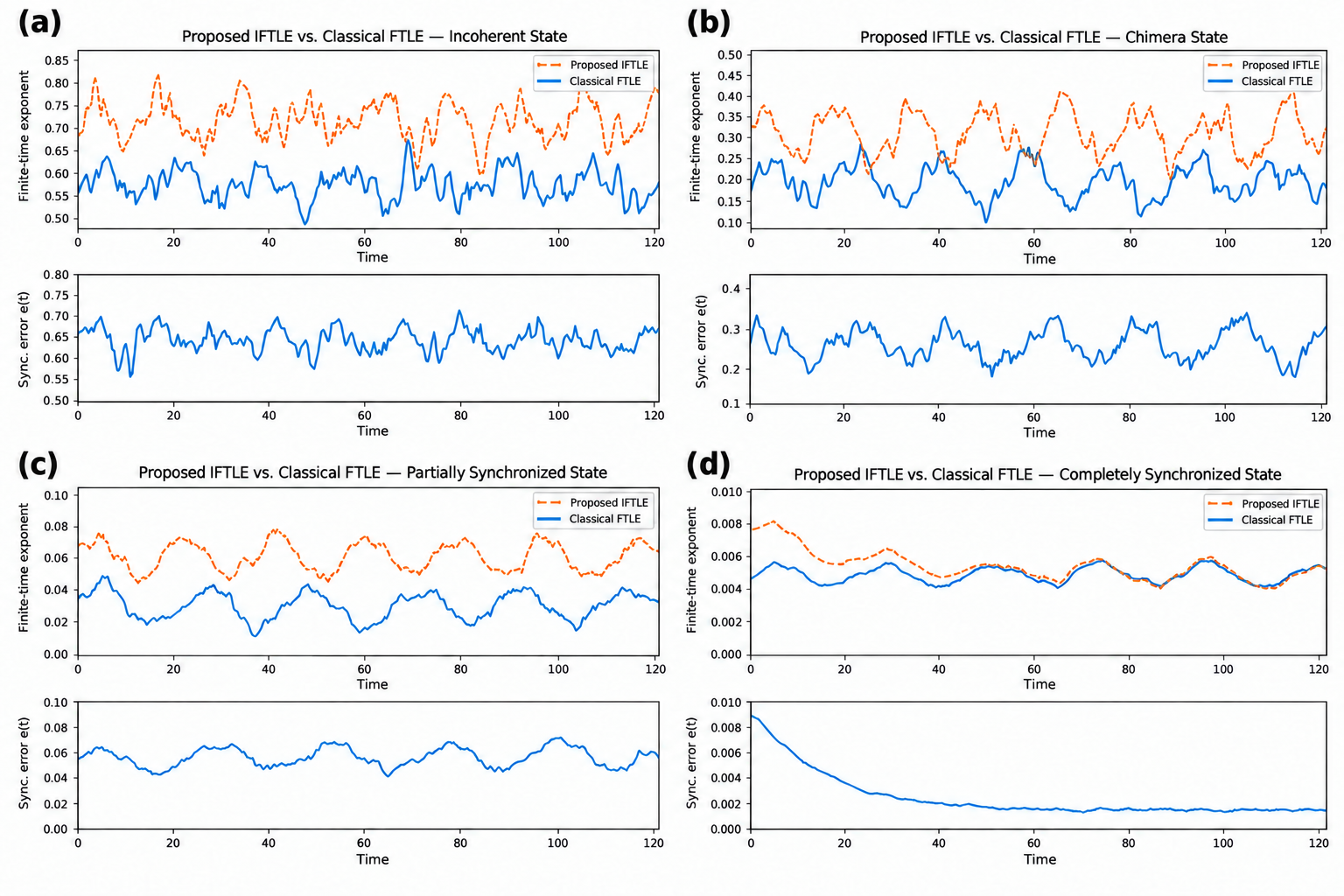}

\caption{
State-resolved comparison of the proposed IFTLE and the  classical FTLE showing the effectiveness of the former
for (a) the incoherent state, (b) the chimera state,
(c) the partially synchronized state, and
(d) the completely synchronized state.
In each panel, the upper subplot shows proposed IFTLE and the classical FTLE, while the lower subplot shows the corresponding synchronization
error $e(t)$ justifying our proposal.
}
\label{fig:appendix_state_comparison}
\end{figure*}

For completeness, we compare the series of values of proposed IFTLE with that of the  classical FTLE 
separately for the incoherent, chimera, partially synchronized, and
completely synchronized dynamical states. We obtain results substantiating our objective that the synchronization error or the deformation due to non-synchrony, hitherto unattended, effectively contribute to the characterization of the underlying collective phenomena.The results are depicted in Fig.\ref{fig:appendix_state_comparison}(a-d)for coherent, chimera, partial synchronization and complete synchronization behaviours, respectively. The upper subplot
shows the temporal evolution of the proposed IFTLE and the classical FTLE,
while the lower subplot shows the corresponding synchronization error
$e(t)$.

\bibliographystyle{aipnum4-2}
\bibliography{references.bib}

\end{document}